# Light Weight Cryptographic Address Generation (LW-CGA) Using System State Entropy Gathering for IPv6 Based MANETs

**Reshmi T.R[1],\*, Murugan K[2]**

[1] VIT University, Chennai, India
[2] Anna University, Chennai, India
\* The corresponding author, email: reshmi.tr@vit.ac.in

**Abstract:** In IPv6 based MANETs, the neighbor discovery enables nodes to self-configure and communicate with neighbor nodes through autoconfiguration. The Stateless address autoconfiguration (SLAAC) has proven to face several security issues. Even though the Secure Neighbor Discovery (SeND) uses Cryptographically Generated Addresses (CGA) to address these issues, it creates other concerns such as need for CA to authenticate hosts, exposure to CPU exhaustion attacks and high computational intensity. These issues are major concern for MANETs as it possesses limited bandwidth and processing power. The paper proposes empirically strong Light Weight Cryptographic Address Generation (LW-CGA) using entropy gathered from system states. Even the system users cannot monitor these system states; hence LW-CGA provides high security with minimal computational complexity and proves to be more suitable for MANETs. The LW-CGA and SeND are implemented and tested to study the performances. The evaluation shows that LW-CGA with good runtime throughput takes minimal address generation latency.

## I. INTRODUCTION

Mobile Ad-hoc Networks (MANETs) are infrastructure-less networks with self-configured hosts communicated via multi-hop communications. IPv6 autoconfiguration was designed to automatically attach a new node to a network and obtain information needed for connectivity. The main goal of autoconfiguration is to have the entire process occur automatically without human interaction, which eases the formation of spontaneous MANETs. Generally key exchange schemes are used to ensure secured communication in MANETs. But these automatic key exchanges can occur only between hosts with established IPv6 addresses. So IPsec is incapable of performing an automatic key exchange, and secure auAtoconfiguration process.

IPv6 uses a hierarchical addressing scheme for the ease of address management. The IPv6 provides an active network interface with a default IPv6 address called the link-local address. This address is fully functional within the local segment and is not routed by routers. Hosts use this address to communicate with other hosts in same network. The 128-bit link local address consists of two 64-bit portions:

a special link-local prefix (FE80:: /10) and a MAC address derived Extended Unique Identifier (EUI-64). The 48-bit MAC is first divided into two 24-bit halves and is filled with 16-bit FFFE in the middle. The 7th bit of the interface identifier (IID) signifies whether the address is global (0) or local (1). The 8th bit is called the group bit; and it signifies whether the address is unicast (0) or multicast (1). The EUI-64 based address generation in Stateless Address Autoconfiguration (SLAAC) [2] is shown in figure 1.

When a node generates a tentative link-local address, it is checked for duplication on the subnet by the Neighbor Discovery Protocol (NDP) [1] assisted process called Duplicate Address Detection (DAD). Each network interface card (NIC) has multiple valid IPv6 addresses such as link-local, assigned unicast, solicited-node multicast, and all-nodes multicast addresses etc. The interfaces are configured with the all-nodes and solicited-node multicast group addresses to perform DAD. During DAD process, the Neighbor Solicitation (NS) messages are sent to the solicited-node multicast address. A Neighbor Advertisement (NA) message allows two nodes to detect the use of the same addresses on the network. When a duplicate is detected, the node with a permanent address, sends back an NA message to the requestor's solicited node multicast address stating the collision. The node then performs regeneration of addresses and DAD up to two more times, after which a warning is written to the system log and the interface attempting to autoconfigure is disabled. If no duplication is detected, the address is considered unique and is assigned as permanent address.

The link-local address is used as an initial default address to retrieve the global and other network-prefix addresses. In a network, when a node receives a Router Advertisement (RA) (unsolicited or as a reply for Router Solicitation (RS)), it creates an IPv6 address appropriate for the advertised network prefix. These addresses are created by attaching the advertised prefix to the already derived EUI-64.

The EUI-64 in IPv6 addresses remains stable for each subnet with the same network interface address, even when the client moves across different networks. So the one-to-one mapping of MAC addresses and EUI-64 in IPv6 addresses paves way to many privacy and security issues. When a node continues using the identity of a EUI-64 based link-local address and retrieves other addresses, it will be permanently bounded to the node and hence the mobility pattern of the node can be easily tracked. Beyond creating privacy issues, it also poses as potential source for information leakage. The EUI-64 based addresses in SLAAC are prone to different types of attacks given below.

- Impersonation: The lack of link layer control and spoofed MAC addresses can indulge any node to claim to be the existing member node in a network by generating the genuine member's EUI-64 based address. These attacks can impersonate as both routers and hosts too. The address spoofing of router results in higher damage, as it can imply man-in-the-middle attacks.
- Denial of Service (DoS): Spoofing of DAD replies can introduce DoS attacks as the neighbor nodes consider that the network is undergoing a network merge or partition. Thereby the attackers will never allow genuine member nodes to participate or get services in the network.

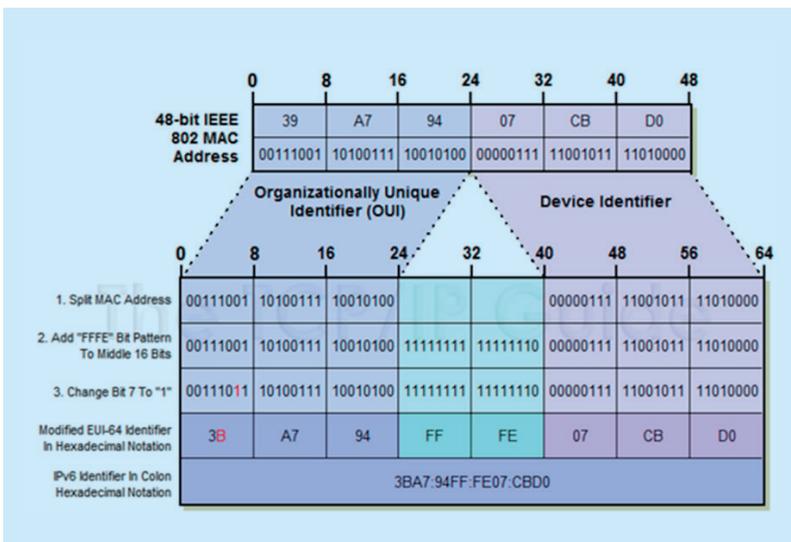

**Fig. 1** *EUI-64 generation in SLAAC*

- Redirection: The methodology of redirection attack is almost similar to the impersonation attack. The attacker misdirects the target node or nodes attempting to connect to the destination nodes by directing to any other unsafe destination to capture the communication packets or interrupt the node communication. Attackers may also announce the change in router address, or network prefixes to interrupt the node communications.

The above three categories of attacks are launched by altering the Internet Control Message Protocol Version 6 (ICMPv6) messages exchanged during autoconfiguration. The details of the ICMPv6 message exchanges are given in table 1.

The various sub classes of attacks launched in the networks during autoconfiguration are also given in table 2. These attacks are launched in both infrastructure and ad-hoc networks with the ICMPv6 messages. The survey [3] states that most of the attacks launched in ad-hoc networks are still not addressed and exist as an open research area. The unique or unpredictable 1nterface identifier based address can reduce the chances of exposure to these attacks. But today there are many solutions available to alter the unique MAC addresses of the devices and hence it is impossible to assure uniqueness of MAC addresses. More over MANETs are dynamic, with frequent merging and partitioning; therefore setting a static policy and expecting it to uphold a secure state is unrealistic. Moreover the Neighbor Discovery (ND) was not designed to deal with security issues and is not suitable for MANETs. Secure Neighbor Discovery Protocol (SeND) [4] addresses few of these issues and it was targeted for infrastructure based networks. As MANETs are infrastructure-less multi-hop communication networks with high resource consumption, there are many technical issues for the implementation of SeND in MANETs.

**Table I** *ICMP message exchanged during autocofiguration*

| Message | Functionality | Source | Destination | Message field options |
|---|---|---|---|---|
| Router solicitation (RS) | Request message for router information | Nodes | All routers | Source link-layer address |
| Router Advertisement (RA) | Response to RS, Advertise autoconfiguration parameters: Default router, on-link prefixes, reachable prefixes, other operation parameters | Routers | Sender of RS/All nodes | Source link-layer address, MTU, Prefixes, Routes, Advertisement interval |
| Neighbor Solicitation (NS) | Request for link-layer address of a target node | Nodes | Target node | Source link-layer address |
| Neighbor Advertisement (NA) | Response to NS, Advertise link-layer address changes | Nodes | Sender of NS/All nodes | Target link-layer address |

**Table II** *Attacks launched during autoconfiguration*

| Attack methodology | Launched attacks | Messages affected | Infrastructure Networks | Ad-hoc Networks |
|---|---|---|---|---|
| NS/NA spoofing | Redirection | NA/NS | Solution known | Solution known |
| Neighbor unreachability detection failure | DoS | NA/NS | Solution known | Solution known |
| DAD DoS | DoS | NA/NS | Solution known | Solution known |
| Malicious router | Redirection | RA/RS | Solution known | Solution unknown |
| Default router kill | Redirection | RA | Solution unknown | Solution unknown |
| Poisoned router | Redirection | RA/RS | Solution unknown | Solution unknown |
| Spoofed redirect | Redirection | NA/NS | Solution known | Solution unknown |
| Bogus on-link prefix | DoS | RA | Solution known | Solution unknown |
| Bogus address config | DoS | RA | Solution known | Solution unknown |
| Parameter spoofing | DoS | RA | Solution known | Solution unknown |
| Replay attack | Redirection | All | Solution known | Solution known |
| Remote ND DoS | DoS | NS | Solution known | Solution known |

**Table III** *ICMPv6 option types in SeND*

| Option Type | Description |
|---|---|
| 1 | Source link-layer address |
| 2 | Target link-layer address |
| 11 | Cryptographically generated address |
| 12 | RSA signature |
| 13 | Timestamp |
| 14 | Nounce |
| 15 | Trust anchor |
| 16 | Certificate |

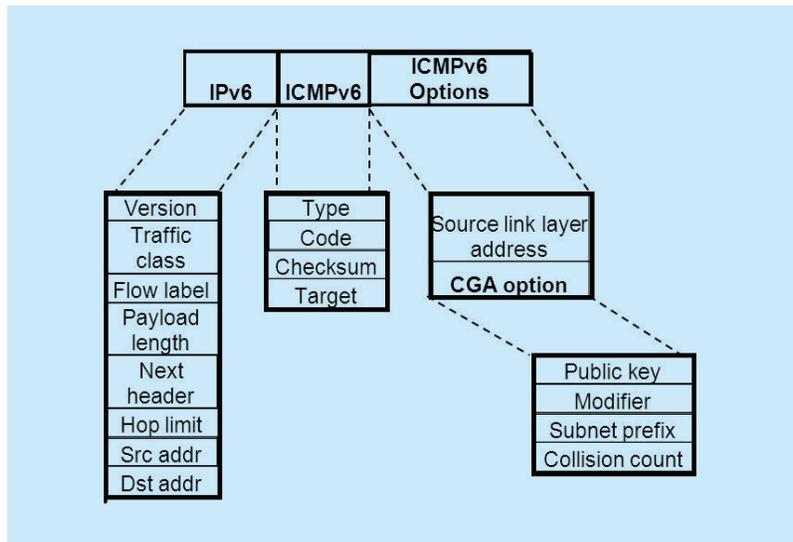

**Fig. 2** *SeND augmented IPv6 packet format*

The proposal is a light weight cryptographic address generation which is an alternate for CGA in SeND, The proposed scheme aims to bring a new address generation technique to adapt to MANET environment and is not targeted to improve the security of the scheme. The motivation of the proposed scheme is to assure:

(1) The IPv6 addresses after CGA generation does not any way relate to MAC addresses and are unique within each subnet for each address in the same network interface of the same client.

(2) The CGA based IPv6 addresses cannot be predicted by the attacker.

(3) The CGA based IPv6 addresses are reconfigured during network merging, partitioning or renumbering.

The paper is organized into five sections. Section 1 discusses the introduction of the paper following the discussion of the existing work in Section 2. The Section 3 discusses the proposed work and the working of the scheme. Section 4 discusses the experimental setup and evaluations. Section 5 discusses the conclusion and the future works.

## II. BACKGROUND

The SeND uses an asymmetric cryptography to enforce authentication and integrity without changing the zero configuration standard of the ND protocol. SeND encodes its ICMPv6 messages [6] by using few new option types that are not used in the regular ND messages. The table 3 lists the new ICMPv6 option types used in SeND. RSA keys and Cryptographically Generated Address (CGA) [5] are used to ensure authenticity in SeND. CGA alike regular IPv6 address has two 64-bit portions.

The SeND augmented IPv6 packets with the different fields of CGA options is given in figure 2.

The first 64 bits represents the network prefix and the second represents the IID, which is derived using the SeND specific CGA generation process. The augmented option fields of CGA in SeND includes Public Key, 128-bit-modifier, 64-bit subnet prefix of the address and 8-bit collision count. The SeND uses two hash functions using the parameters given in the CGA option field. The HASH1 and HASH2 functions play a major role in CGA generation. The generations of CGAs involve determining the public key of the owner address, selecting the appropriate security-level (Sec) ranging from 0 to 7 and generation of a random 128 bit modifier. Then it is subjected to SHA-1 hashing and is looped continuously with various values assigned to the modifier until $16 \times Sec$ leftmost-bits of HASH2 equals zero. This final modifier value is again used along with the various other CGA parameters as the input to the HASH1. HASH1 is the leftmost 64-bit of the SHA-1 hash function. It concatenates the modifier, subnet, collision count and the public key fields of CGA options. HASH1 resultant is modified and used as the IID in CGA. The seventh bit of HASH1 is modified to represent whether the address is global or local. The eighth bit represents whether the address represents unicast or multicast groups. TheHASH1 uses a hash extension technique using the Security Parameter (Sec) to decide the strength of the hash function. In CGA Sec is used as the first 3 bits of the IID. The CGA generation process is represented in figure 3.

During the verification process, the CGA options in the ICMPv6 packets are first extracted to calculate the HASH1 and HASH2 values. At first the bit sequence (excluding seventh and eighth bit) of the calculated HASH1 is compared with the IID. The SeND daemon then compares the 16 x Sec leftmost bits of HASH2 to zero. If any of these comparisons fails, the packet processing is stopped and the packet is discarded. Following these verification the digital signatures are verified. The RSA signature method helps to prove whether the public key corresponds to the private key of the packet sender. This binds the CGA and the key pair of the same origin. Even though authentication is well addressed by CGA and RSA, authorization is still a challenging issue in SeND. Hence this scheme is prone to rouge router attacks like poisoned router, malicious router, default router kill etc. (discussed in table 2).

Both packet size and the computational intensity for CGA generation in SeND is a major concern for MANETs. In general, MANETs are at risk as it would be easy for an attacker to flood the network with large, expensive-to-process packets, consuming bandwidth, resources and energy, resulting in limited responsiveness of the nodes. Due to the frequent exchange of packets during merging and partitioning of MANETs, SeND employs reconfiguration of nodes that would greatly increase the resource consumption of nodes.

A SeND context is an internal data structure that stores information about the interface, options, prefixes, and keys used by packets. So for every packet that fails or pass the verification process adds a context to the stored information, and hence adds unnecessary computation. The computational cost of a CGA generation, with a non-zero Sec value, increases the address configuration time. Even though there are many modifications suggested in SeND, these schemes couldn't overcome the issues faced by the existing schemes. The privacy addresses generated in a scheme proposed by Narten et al. [7] used the Message Digest (MD5) on the message stored in a stable storage to retrieve the addresses. But these addresses can be easily predicted by tracking the file usage history. Bos et al. [8] proposed an analyzed optimal CGA generation, but the scheme is for infrastructure based networks and hence is not adopted for MANETs. Jiang [9] introduced an interactive method of address generation with DHCPv6, but the dynamic nature of MANETs restricted assigning dedicated services and interactive communication packets.

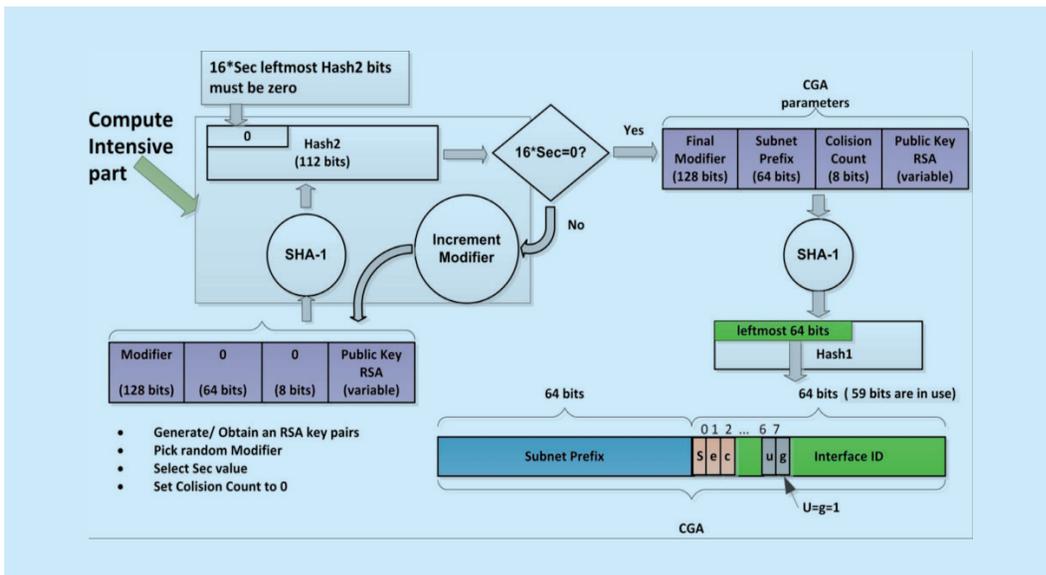

**Fig. 3** *CGA generation*

The extended SeND [10] and windows based SeND [11], were not implemented and standardized because of the limitation of applicability. The cryptographic algorithms of SeND are extended or modified in many schemes [12-19] by replacing RSA with ECC and newly designed cryptographic algorithms to reduce the computational complexity of authentication. But as MANETs are provided with low resources, these schemes may drain the resources and hence are not applicable for MANETs. SAVI [20] scheme is a link-layer authentication protocol scheme which ensures the authenticity of the packets exchanged during autoconfiguration. But it fails to assure protection against other ND vulnerabilities. The surveys [21-23] on the existing schemes emphasize the requirement for a light weight and empirically strong CGA generation for resource constrained MANETs and is the motivation for the proposed work.

## III. PROPOSED WORK

The SeND protocol designed to overcome the security threats during autoconfiguration has proven to face security and technical issues in MANETs (discussed in Section 2). The paper proposes an unpredictable random number based address generation using internal system states, for ensuring security of autoconfiguration with minimum resource consumption. The internal system states based address generation is an inspirational proposal of Hardware Volatile Entropy Gathering and Expansion (HAVEGE) [24]. Pseudo random numbers exhibiting high degree of randomness are needed for highly secured cryptographic algorithms.

A heuristic algorithm that relies on entropy gathered from unpredictable system events is proposed. This entropy gathering technique is used in pseudorandom number generator to generate the IID during autoconfiguration. The algorithm implements a hardware clock cycle counter to gather the entropy from the system states of the nodes. The system states are hardware mechanisms that improve performance of caches, branch predictors and external devices. These system states are not architectural, but volatile and cannot be directly monitored by the user and hence are used as source for highly random pseudorandom numbers. Since the system state includes thousands of internal volatile hardware states, it is impossible for the user to reproduce the generated bit sequences. Any attempt to indirectly gather the bit sequence triggers the internal state of the system and therefore, reproducing the bit sequences is virtually impossible. Although the proposed address generation method is a light weight scheme with fewer computations, it has proven compliance to all the security standards discussed in NIST statistical test suite [25]. The internal system states based address generation called as Light Weight Cryptographic Address Generation (LW-CGA) is a novel light weight autoconfiguration scheme designed to ensure high security and adaptability in MANETs.

### 3.1 Light weight cryptographic address generation (LW-CGA)

The LW-CGA algorithm uses a heuristic algorithm to collect entropy of system states and generate a sequence of random number bits as IID. A hardware clock counter is used to gather the sequence of uncertainty of the system states. The entropy gathered from the system states are generated from the instruction cache and branch prediction structures. A function to read the hardware clock called *CLKREAD( )* is used and it verifies the difference with last read values. The counter *INTERRUPTCOUNT* is incremented by function *CLKREAD( )* when the difference of the clock read is higher than a threshold level MAX indicating an interrupt between two successive reads. Throughout the algorithm, *CLKREAD( )* is called several times and the resultant is combined by XOR and shifts in an array *ENTROPY[ ]*. Since the entropy in the least significant bits (LSB) of the hardware clock counter is more than the most significant (MSB), the read value is combined with circular shift of the previously accumulated data for even diffusion of entropy throughout the array *ENTROPY [ ]*. The entropy collection loop runs until the *INTERRUPTCOUNT* reaches a predefined threshold. *BUFFERSIZE* is the size of the table used to gather the values of the hardware clock counter. At an instance the content of the *ENTROPY[ ]* is saved and reinitialized to zero. The saved *ENTROPY[ ]* is combined with a simple pseudorandom number generator. A *DYNAMTABLE[ ]* is used to record the two instants of updates in the single *ENTROPY[ ]* table. *SCROLL[ ]* is a memory table which is twice the size of the L1 Cache (discussed in Section 3.2.1 and is assumed in power of 2). The *SCROLL[ ]* stores the resultant empirically strong random number which is assigned as the IID of the CGA addresses. The random bit sequence from the internal micro architectural status of a system and integration to the link local address makes the addresses unpredictable. The figure 4 shows the pseudocode of LW-CGA algorithm.

The algorithmic steps are given below.

*Step 1:* The function *CLKREAD()* is defined to read and return the hardware clock counter value of the node.

*Step 2:* Two concurrent *SCROLL* operations are performed in parallel in a table of 4B. The table 1s twice the size of L1 data cache. If the *SCROLL* is random, then the probability of a hit in the cache is very close to 1/2 on each data reading from the table.

*Step 3:* Two data dependent tests are introduced on iterations of *SCROLL* to make its behavior depend on branch prediction information. For both branches, the probability of the branch being taken is 1/2 if the content of the table 1s random.

*Step 4:* The *SCROLL* is iterated to the number of unrolled steps (y) for data reads from instruction cache. This maximizes the number of instruction blocks (and associated branch prediction information) extracted from the instruction cache on each operating system interrupts.

*Step 5:* The two distinct data read (Steps 3 and 4) acquired during *SCROLL* performs Exclusive-OR (XOR) in memory table. If memory table 1s directly read for random number generation, then an observer can follow up the *SCROLL* for a while and try to guess the partial content of the table. XOR is used to hide the content of the *SCROLL* table from any possible observer.

*Step 6:* The 64 bit Least Significant Bits (LSB) extracted from random bits of the output of the algorithm are used as IID for CGA generation.

### 3.1.1 System states for random number generation

The system states of the nodes are unpredictable 5olatile hardware states and are activated by the instruction cache and branch predictors. Any node with a processor can implement the algorithm and collect the system states. The

```
initialize int ENTROPY[ ]
initialize int N;
initialize int SCROLL[2*CACHESIZE];                      //initialized during entropy
initialize register int pt, PT, PT2;                     //initialized during entropy
initialize register int i;
while (INTERRUPTCOUNT < MAX) do
        for(a=1; a<x; a++)                  //x is initialized based on instruction cache and branch
        if (N==0) N++; else N --;
            ENTROPY[X] = (ENTROPY[X] << 5) ^ (ENTROPY[X] >> 27) ^ CLKREAD() ^
                        (ENTROPY [(X+1) & (BUFFERSIZE -1)] >> 31 );
            X = (X+1) & (BUFFERSIZE -1);
        end if
end while
For(n=1; n<y; n++)                                //y is initialized based on the processor
    if (pt & 8*CACHESIZE) ;
    if (pt & 16*CACHESIZE) ;
    PT = pt & (2*CACHESIZE-1); pt = SCROLL [PT];
    PT2 = SCROLL [(PT2 & CACHESIZE-1) ^ ((PT ^ CACHESIZE) & CACHESIZE)];
    RESULT[i] = PT2 ^ pt;
    Threshold = ((T << 7) CLKREAD ()) ^ (T >> 25);
    pt = pt ^ T;
    SCROLL [PT] = pt;
     i++;
extract 64-bt LSB
initialize and generate CGA
End
```

**Fig. 4** *Algorithm of CGA generation in LW-CGA*

volatile states of the nodes are influenced by the following components.

i. DATA Translation Look-aside Buffer (TLB): The TLB buffer is a table consisting of information of the pages in the memory that are accessed by the processor. Each entry of a page in the TLB possesses about 129 states which invoke system states. The LW-CGA is constructed with a loop that access 128 pages of memory.

ii. L1 Data Cache: The L1 cache also known as the primary cache memory in the processor core possesses information of the latest data used by a processor. In the LW-CGA, the cache lines of the L1 data cache are assumed to be in one of the seven possible states. This data is usually mapped in the SCROLL table between any one of the available 32 byte block .The L1 cache is selected as one of the 7512 possible states. The processor might additionally possess L2 and L3 caches which are not considered in this implementation.

iii. L1 Instruction Cache: The L1 instruction cache consists of 256 sets in which, each set possesses 7 possible states. The LW-CGA is made of a loop body that performs self-modifying paces over the level 1 instruction cache.

3.1.2 Regeneration options of LW-CGA

The LW-CGA includes several options to regenerate the CGA, resulting in different IPv6 addresses. These regenerations are explicitly prompted by a function call *REGENERATE( )* in the daemon. The various options provided by the LW-CGA regeneration are listed below.

i. Regeneration on updated RA and NA: The RA and NA message contains new prefix advertisements or updates to indicate prefix-change or MANET merging or partitioning. So whenever an advertisement with new prefix is received the daemon calls *REGENERATE( )*

ii. Regeneration on interface status change: The option enables a new IID generation when a node enables a network interface. So when an interface is enabled state it will be configured with a new address even if the network-prefix remains the same. By default as in existing schemes, each node with multiple interfaces will be configured with different IIDs.

iii. Regeneration on user choice: The CGA generation can be explicitly invoked by the user based on his/her choice. This feature is added for testing purpose and entropy calculation on each generation.

iv. Regeneration at regular intervals: The regeneration at regular interval is an option included for maintenance and security purposes. The option has been included to add the renumbering features.

## IV. EXPERIMENTAL EVALUATIONS

The LW-CGA and SeND autoconfiguration schemes are implemented in real networks for experimental evaluations. The flow chart of LW-CGA and SeND implementation is shown in figure 5.

### 4.1 Experimental set-up

The Linux kernel version of 2.6.34 is used in nodes for the implementation of the LW-CGA and SeND autoconfiguration schemes. The experimental analysis can be interrupted by the scheduler as its time is accounted for the measurements. Hence single mode Linux kernels are used for experimentations. The SLAAC implementation of IPv6 is modified for the implementations. The modified kernel provides several sys-controls which can be read and written to and by user-space programs, controlling the operation of IPv6 privacy extensions. SeND-CGA project developed and maintained by Huawei Technologies Corp. and BUPT (Beijing University of Post and Telecommunications) [26], is implemented as a modified kernel and invoked with a daemon. The LW-CGA is an extension of HAVEGE [24] and the same is implemented as a modified kernel. The kernel sys-controls are adjusted to characterize MANETs. No changes pertaining to IPv6, ICMPv6, and NDv6 parameters were made, so as to preserve the other network layer features. SeND is compiled with the minimal options required to compile on Linux and is used for the testing.

The LW-CGA as like SeND uses the RSA based key (1024 bits) exchange scheme during the first CGA generation. But during regeneration of CGA (caused by node mobility or change in network gateway), the keys are not regenerated but verified for authenticity. In SeND scheme, the HASH1 alone is recomputed during CGA regeneration. The SeND evaluation of Sec values higher than 1are not validated as it is proven to be impractical [23]. The following parameters are used for evaluations.

- IID generation time is the total time duration for the generation of public key, their verification and computation of CGA followed by IID configuration
- IID regeneration time is the time taken for the IID regeneration by CGA regeneration without key generation. The CGA regeneration of SeND requires a HASH1 calculation alone, whereas only a random number generation in LW-CGA.
- Entropy is the measure of randomness in a closed system. The entropy of a random variable $X$ with probabilities $p_i, ..., p_n$ is defined as

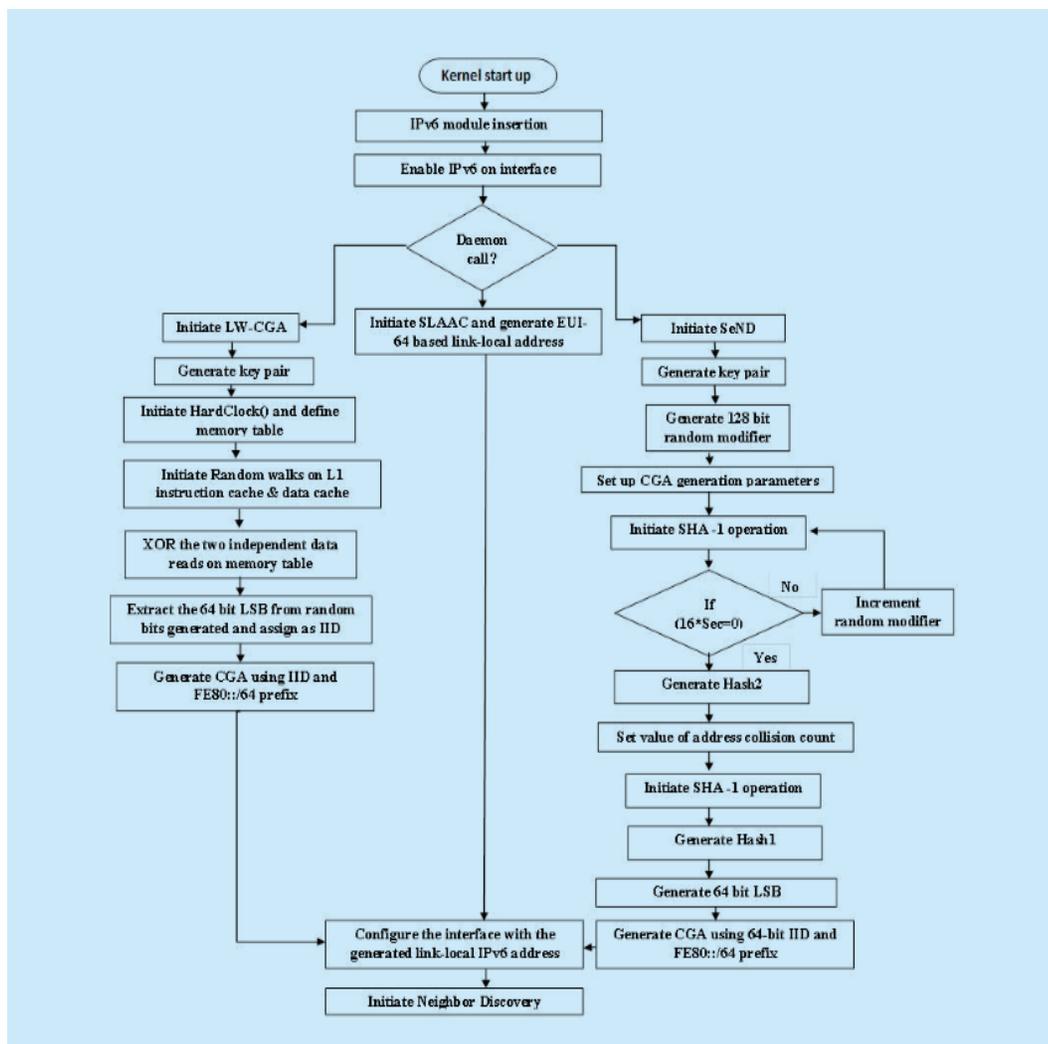

**Fig. 5** *Flow Chart of LW-CGA and SeND Implementations*

China Communications • September 2017　　　　　　　　　　　　　　　　　　　　　　　　　　　　　　　　　　　　　　　　　122

$$H(X) = \sum_{i=1}^{n} p_i \log p_i \qquad (1)$$

The NIST Test Suite [25] is a statistical package with 15 tests to test the randomness or entropy of the binary sequences generated by the algorithms. We have conducted NIST tests to test the randomness of binary sequences generated by SHA-1 in SeND and random number generator in LW-CGA. These tests focus on a variety of non-randomness that could exist in a sequence. Various statistical tests were applied to the bit sequence generated by LW-CGA and SeND to compare and evaluate the randomness. The randomness of bit sequences was characterized and described in terms of probability. These statistical tests were formulated to test a specific null hypothesis (H0). The null hypothesis was the sequence being tested is random. Associated with this null hypothesis is the alternative hypothesis (Ha), for which the sequence is not random. For each test, the decision declares the acceptance or rejection of null hypothesis, i.e., whether the sequence generated is truly random or not. In each test, a relevant randomness statistic was chosen to determine the acceptance or rejection of the null hypothesis

A mathematical method was used to determine the theoretical reference distribution of this statistic under null hypothesis. A critical value was selected from this reference distribution (typically, this value is "far out" in the tails of the distribution say, out at the 99 % point). During the tests, the statistic value computed on the bit sequence was compared to the critical value. If the test statistic value was higher than the critical value, the null hypothesis was rejected. Otherwise, the null hypothesis was accepted. For these tests, the probability of randomness determined as P-value was compared with the derived significant level (α). If P-value ≥α, then the bit sequence was considered random and if it was not so, then it was considered non-random. Typically, α was chosen in the range 0.001 and 0.01.

Fifteen tests of NIST Statistical Test Suite have significant purpose to confirm the randomness of the bit sequence. The Block and Frequency Monobit tests determine whether the number of ones and zeros in a sequence are approximately the same as expected for a truly random sequence. The Binary Matrix Rank Test checks for linear dependence among fixed length substrings of the original sequence. The Longest Run Test determine whether the length of the longest run of ones within the tested sequence is consistent with the length of the longest run of ones as expected in a random sequence. The Run Test determines whether the oscillation between ones and zeros of various lengths is as expected for a random sequence. Spectral Discrete Fourier Test (DFT) detects repetitive patterns in the tested sequence to indicate a deviation from the assumption of randomness. Non-Overlapping Template Matching Test checks whether too many occurrences of a given non-periodic pattern occur in tested sequence. Overlapping Template Matching Test determines the number of occurrences of pre-specified target strings. It uses the same m bit window as in Non-Overlapping Template Matching Test, to search for a specific m bit pattern (B). The difference is that when the pattern is found, the window slides only one bit before resuming the search.

The Maurer's Universal Test detects whether the sequence is significantly compressed without loss. The Linear Complexity Test determines whether the sequence is complex enough to be considered random. The Serial Test determines whether the number of occurrences of the 2m m bit overlapping patterns is approximately the same as expected for a random sequence. Approximate Entropy Test compares the frequency of overlapping blocks of two consecutive lengths (m and m+1) against the expected result for a random sequence. The Cumulative Sum (cusum) Test determines whether the cumulative sum of the partial sequences in the sequence is too large or too small relative to the expected behavior of cumulative sum for random sequences. The Random Excursion Test determines whether the number of visits to a particular state within a cycle deviates from that expected for a random sequence. The Random Excursion Variant Test detects deviations from the expected number of visits to various states in random scroll.

### 4.2 Experimental results

The CGA generation time of SeND and LW-CGA is measured with an internal counter clock. The experimental results of more than 100 samples have been averaged to validate the analysis. The result shown in table 4 concludes that the LW-CGA takes less time when compared to SeND. The key generation time is almost the same for both the schemes, but the verification and CGA generation time is high in SeND which contributes to extra time consumption.

The CGA generation time for Sec=1 is greater than Sec=0 of SeND. The difference in time is to generate the modifier that requires further time for calculating HASH2 values until the 16 x Sec leftmost bits are zeros. Hence at Sec=1 the key generation followed by verification, HASH1 and HASH2 computation for CGA generation time contribute to the total IID generation time. At Sec=0, the key generation time, verification time and HASH1 computation time for CGA generation time only contribute to the total IID generation time. The HASH1 and HASH2 computation involves the use of public key as a parameter; hence these computations are sequential and are interdependent (discussed in Section 2). The LW-CGA uses a key exchange scheme for verification and validation of the keys generated. The CGA generation scheme in LW-CGA is not dependent on the keys and involves random bit generations from entropy gathered from system states. LW-CGA scheme uses a less computational intensive method for CGA generation and hence the time taken for the IID generation is comparatively very less compared to the SeND.

The CGA regeneration time of SeND at Sec=0 and Sec=1 does not involve key exchanges and verifications, but computation of HASH1. The LW-CGA just recertifies the authenticity of the keys and regenerates the random number. The results conclude that the time required for CGA generation in LW-CGA is very less when compared to SeND.

The security of the schemes can be evaluated with the entropy of the algorithms. The security flaws of SHA-1 have invoked many on-going works in the CGA and SeND maintenance working group in IETF. These have issued the recommendation of SHA-2, SHA-256 etc. But these schemes are highly computational intensive and fail in resource constrained environments. The SHA-1 in SeND and random number generator in LW-CGA are tested using the NIST statistical test suite [25]. The recommended statistical tests are done to evaluate the performance in security algorithms. For a p-value $\geq 0.001$, the sequence is considered to be random with a confidence of 99.9% or else nonrandom with a confidence of 99.9%. The analysis shows that random number generator of LW-CGA performs well in cryptanalysis.

## V. CONCLUSION AND FUTURE WORKS

SeND uses RSA and SHA-1 implementation for ensuring privacy enabled autoconfiguration. The generation time and the computational intensity of CGA make SeND implementation impractical for MANETs. Moreover the improved security level of SeND varying from 0 to 7 worsens the computational intensity and battery consumption in wireless devices. The paper proposed a light weight cryptographic scheme called LW-CGA that ensures high security with minimal and faster computations. The random number generations in the scheme uses an entropy gathering algorithm from the system states and are independent of the keys. The bit sequences generated by the algorithm are impossible for guessing and cannot be even monitored by the system users. Hence the scheme is highly random and assures security. It's proven that the usages of LW-CGA satisfies all the characteristic requirements of cryptographic algorithms and are more suitable for constrained devices like PDA and Tablet PC. The schemes are evaluated with real time implementation to study CGA generation using SHA-1 in SeND and random number generation in LW-CGA. The experimental results show significantly reduced address generation time while using LW-CGA. The NIST statistical tests demonstrate that a higher security is assured with LW-CGA without the need for a high cost algorithm. Future works of the authors mainly focus on more light weight techniques for privacy enabled autoconfiguration in MANETs.

Table IV  *IID generation time of SeND (at Sec=0 and Sec=1) and LW-CGA*

| Time Taken | SeND (Sec =0) (in Seconds) | SeND (Sec =1) (in Seconds) | LW-CGA (in Seconds) |
|---|---|---|---|
| Key generation time | 0.01629 | 0.02394 | 0.01656 |
| Key verification time | 0.00045 | 0.00132 | 0.00042 |
| CGA generation time | 0.62437 | 1.89843 | 0.21276 |
| IID generation time | 0.64111 | 1.92369 | 0.22898 |

Table V  *IID regeneration time of SeND (at Sec=0 and Sec=1) and LW-CGA*

| Time Taken | SeND (Sec =0) (in Seconds) | SeND (Sec =1) (in Seconds) | LW-CGA (in Seconds) |
|---|---|---|---|
| IID generation time | 0.65489 | 2.4591 | 0.27832 |

**Table VI** *NIST statistical tests on SeND and LW-CGA*

| NIST Statistical Tests* | p-value of SeND | p-value of LW-CGA |
|---|---|---|
| Frequency | 0.604458 | 0.859684 |
| Block Frequency ($m = 128$) | 0.091517 | 0.483376 |
| Cusum-Forward | 0.451231 | 0.696011 |
| Cusum-Reverse | 0.550134 | 0.574764 |
| Runs | 0.309757 | 0.360215 |
| Long Runs of Ones | 0.657812 | 0.908779 |
| Rank | 0.577829 | 0.682148 |
| Spectral DFT | 0.163062 | 0.220375 |
| Non Overlapping Templates ($m = 9$, $B = 000000001$) | 0.496601 | 0.635352 |
| Overlapping Templates ($m = 9$) | 0.339426 | 0.879858 |
| Universal | 0.411079 | 0.800156 |
| Approximate Entropy ($m = 10$) | 0.982885 | 0.998663 |
| Random Excursions ($x = +1$) | 0.000000 | 0.277462 |
| Random Excursions Variant ($x = -1$) | 0.000000 | 0.600720 |
| Linear Complexity ($M = 500$) | 0.309412 | 0.637938 |
| Serial ($m = 16$, $2m\nabla\Psi$) | 0.760793 | 0.716996 |

**Biographies**

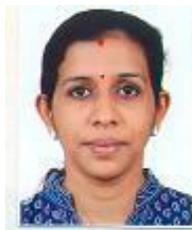

*Reshmi T.R,* received her Ph.D from Anna University, Chennai. She is currently working as Assistant Professor (Senior) in VIT University, Chennai. She had been certified as the IPv6 Forum Certified Engineer (Silver). She has a good publication record in reputed journals and conferences. Her area of interest includes Wireless Network security, SDN and IPv6 security and service applications.

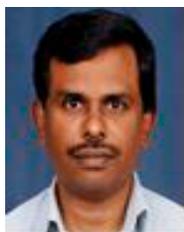

*Murugan K,* received his Ph.D from Anna University, Chennai. He completed his Masters in Computer Science and Engineering, at National Institute of Technology, Tamilnadu. He is currently working as Professor at Ramanujan Computing Centre, Anna University, Chennai, India. He had published about 100 highly reputed journals and 300 conference papers. He is a life member of IETE, ISTE, and CSI. His area of interest includes Wireless Networks, Internet of Things (IoT), Cognitive networks etc.